\documentclass[a4paper,aps,showkeys,nofootinbib,preprintnumbers,superscriptaddress,amsmath,amssymb,amsfonts,floatfix]{revtex4}
\usepackage{graphicx}
\graphicspath{{./images/}}
\usepackage{amsmath}
\usepackage[latin1]{inputenc}
\usepackage[english]{babel}
\usepackage[T1]{fontenc}
\usepackage{amssymb}
\usepackage{amsfonts}
\usepackage{epsfig}
\usepackage{colordvi}
\usepackage{psfrag}
\usepackage{color}
\usepackage{dcolumn}
\usepackage{multirow}
\usepackage{hyperref}
\usepackage{epstopdf}
\usepackage{times}
\hypersetup{
  colorlinks=true,        
  linkcolor=blue,         
  citecolor=magenta,      
}

\begin{document}
\title{Gravitational self-force and conservative effects: a testing ground for theories of gravity}

\author{Vincenzo Ventriglia}

\date{\today}

\begin{abstract}
Considering extreme-mass-ratio inspirals along with the conservative dynamics of gravitational self-force, we compare viable theories of gravity. In particular, by examining a Schwarzschild background we analyse the self-force-induced corrections to gauge-invariant benchmarks given by the orbital frequency at the ISCO and the spin-precession rate. Moreover, following an established indication of modifications to the equations of motion in extended theories of gravity, we exploit a phenomenological approach -- relying on the variability of the gravitational constant $G$ -- to incorporate these modifications. We find that conservative effects shape up to be a test-bed for theories of gravity, allowing us to contrast General Relativity with competing theories. By examining strong-field constraints, we highlight a wide margin of investigation in the context of LISA Mission.
 \end{abstract}
 \pacs{00000}

\maketitle

\section*{Introduction}
The landmark observation of merging black holes (BHs) by the LIGO-Virgo Collaboration in 2015 \cite{LIGO-Virgo2016} has conclusively brought BHs, inspiraling BH binaries and gravitational waves (GWs) in the realm of physics rather than mathematical speculation, marking the birth of GW astronomy. The LIGO-Virgo discoveries would not have been possible without an accurate model of the inspiral and merger; actually their analysis \cite{LIGO-Virgo2017} concluded that the quality of science extractable from future observations may well be limited not by experimental precision, but by the accuracy of available theoretical models.

One of the next steps in the gravitational astronomy programme is the observation of an inspiral scenario where one of the BHs is much lighter than the other one -- the so-called ``extreme-mass-ratio inspiral'', or EMRI. Such an astrophysical system describes the long-lasting inspiral (from months to a few years) and plunge of a stellar-origin BH, with mass $m\sim 10-60\ M_{\odot}$, into a supermassive BH of mass $M\sim 10^{5}-10^{6}\ M_{\odot}$ located at a galactic centre. The orbits of EMRIs are expected to be generic and highly relativistic; in addition, depending on the chirp mass of the system, the GW signal can stay in the LISA \cite{LISA_newdesign} bandwidth for years. The small object may then spend many cycles in close vicinity of the supermassive BH, with its orbit displaying extreme forms of general-relativistic effects. The intricate GW signal encodes within it a detailed mapping of the spacetime around the supermassive BH, allowing us to (\textit{i}) test its geometry and accurately measure its mass and spin, and (\textit{ii}) confirm whether it is a BH as General Relativity (GR) predicts, thus eliminating or tightly constraining a heap of proposed alternatives to GR: a laboratory for strong-gravity physics.

Nature seems to abound with EMRIs, which are expected to emit GWs in mHz frequencies, impossible to be detected by Earth-borne detectors because of the seismic noise limiting the sensitivity to frequencies above approximately 1 Hz \cite{aLIGO_transient_noise,aLIGO_sensitivity}. Indeed, EMRIs are prime targets for the planned LISA mission \cite{LISA}, whose peak sensitivity will be exactly in the mHz band, and typical estimates for the detection rate for LISA range from a few to hundreds of events per year \cite{Amaro-Seoane2007}.

The desire to maximise the science return drives the theory programme to improve waveform models across the full parameter space relevant to the observation. However, developing templates able to follow the actual waveforms with high accuracy over $10^{4-5}$ cycles is a highly non-trivial issue. In fact, dealing with exquisitely relativistic systems such as EMRIs brings major difficulties: the post-Newtonian expansion is cumbersome, since it expands about flat spacetime and it would converge too slowly; Numerical Relativity, on the other hand, digs into the full non-linear dynamics, but it is also unsuitable, as relatively long numerical simulations are currently only possible for mass ratios up to about 10 -- due to numerical complications introduced by the different length-scales associated with the two BHs. A synergistic approach, however, seeks to interface between these results in intermediate domains, which may not be accessible to either of the approximation methods separately \cite{LeTiec2014}. One can however make use of the fact that in EMRIs the mass ratio $\eta\equiv m/M$ is small, e.g. of order $10^{-4}-10^{-6}$, so one can perform an expansion in this parameter, giving rise to a perturbative approach.

This paper is organised as follows. In Sec. \ref{sec:GSF} we briefly introduce and discuss the concept of gravitational self-force (GSF) in light of recent advances. In Sec. \ref{sec:ETGs} we pause the main discussion in order to motivate and introduce the concept of Extended Theories of Gravity (ETGs); then we deal with the analysis of self-force in scalar-tensor gravity, finally specialising our discussion to the case where both objects in question are BHs. The analysis of conservative effects will then engage us throughout Sec. \ref{sec:ConsEff}, examining these effects first in GR and then moving on to an ``effective'' theory -- relying on a theory-agnostic parametrisation -- that takes into account the corrections induced by ETGs. Finally, in Sec. \ref{sec:S/N} we briefly discuss EMRIs detectability in the context of LISA.
Throughout this work, the use of geometrized units -- in which $G=c=1$ -- is implied, unless otherwise specified.

\section{Gravitational self-force\label{sec:GSF}}

Considering a small, compact object which moves in a curved spacetime, one might ask what path does it follow. Indeed, beyond the test-particle approximation, a real object does perturb the geometry and does feel its own field. The physical perturbation due to the particle -- a retarded solution of the linearised Einstein equations, $h_{\mu\nu}^{\mathrm{ret}}$ -- is singular at the location of the particle; hence, stating that the particle follows a geodesic of $g'_{\mu\nu}=g_{\mu\nu}+h_{\mu\nu}^{\mathrm{ret}}$, would be physically meaningless.

In 1996 two groups (Mino, Sasaki, Tanaka \cite{MiSaTa}, and Quinn, Wald \cite{QuWa}) obtained independent derivations of the GSF, since then known as the MiSaTaQuWa equation. Their work was crucially inspired by the classical analyses of the electromagnetic self-force problem, both in flat (Dirac, 1938 \cite{Dirac1938}) and curved (DeWitt and Brehme, 1960 \cite{DeWitt1960}) spacetimes. At any spacetime point $x$, the perturbation can be written as $h_{\mu\nu}^{\mathrm{ret}}=h_{\mu\nu}^{\mathrm{dir}}+h_{\mu\nu}^{\mathrm{tail}}$, \textit{i.e.} a direct contribution from the intersection of the past light-cone of $x$ with the worldline $\Gamma$ of the particle, and a tail contribution arising from the part of $\Gamma$ inside the light-cone -- whose occurrence is a well-known feature of the wave equation in a curved spacetime, with waves being scattered off spacetime curvature. Detweiler and Whiting \cite{DetWhi2003} provided a more compelling decomposition of the retarded field, $h_{\mu\nu}^{\mathrm{ret}}=h_{\mu\nu}^{\mathrm{S}}+h_{\mu\nu}^{\mathrm{R}}$, whose advantage is in bolstering the interpretation of the fields as self-fields. In fact, the S-field, as well as the direct one, exhibits a $1/r$, Coulomb-like divergence at the particle; however, unlike the direct field, it satisfies the inhomogeneous linearised Einstein equation $G_{\mu\nu}\left[h^{\mathrm{S}1}\right]=-16\pi T_{\mu\nu}^{1}$. Similarly, the R-field includes the backscattered radiation in the tail, being also a vacuum solution of the homogeneous wave equation $G_{\mu\nu}\left[h^{\mathrm{R}1}\right]=0$ smooth on the worldline. In terms of the R-field, the MiSaTaQuWa equation can be written as 
\begin{align}
    a^\mu \equiv & \frac{D^{2}z^{\mu}}{\mathrm{d}\tau^{2}} \nonumber \\
    = & -\frac{\epsilon}{2} P^{\mu\nu}\left(2\nabla_{\beta}h_{\nu\alpha}^{\mathrm{R}}-\nabla_{\nu}h_{\alpha\beta}^{\mathrm{R}}\right)u^{\alpha}u^{\beta}+O\left(\epsilon^{2}\right), \label{eq:MiSaTaQuWa_R}
\end{align}
where $D/\mathrm{d}\tau\equiv u^{\mu}\nabla_{\mu}$ is the directional covariant derivative, $P^{\mu\nu}\equiv g^{\mu\nu}+u^{\mu}u^{\nu}$ projects orthogonally to the worldline and $\epsilon$ is the spatial geodesic distance from $x$ to $\Gamma$.

The idea that the GSF can be expressed as a back-reaction from a smooth vacuum perturbation leads to an interesting re-interpretation of the GSF effect: \textit{the particle effectively moves freely along a geodesic of a smooth perturbed spacetime with metric}
\begin{equation}
    g_{\mu\nu}+h_{\mu\nu}^{\mathrm{R}}\equiv\tilde{g}_{\mu\nu}. \label{eq:tildeg}
\end{equation}
This is very welcome, because it is more in the spirit of GR equivalence principle and teaches us that self-accelerated motion in $g_{\mu\nu}$ is equivalent to geodesic motion in the effective metric $\tilde{g}_{\mu\nu}$. For a recent and pedagogical introduction to the topic, see, \textit{e.g.}, \cite{BarPou2018}; while for a more technical review, we refer the reader to \cite{PoiPouVeg2011}.

A crucial point in the analysis of predictions of the GSF theory is the acknowledgement that different pieces in the MiSaTaQuWa equation \eqref{eq:MiSaTaQuWa_R} contribute to different effects. By this we mean that it can be written as a sum of time-symmetric and time-antisymmetric parts, thus yielding $F_{\mathrm{self}}^{\mu}=ma^\mu=F_{\mathrm{cons}}^{\mu}+F_{\mathrm{diss}}^{\mu}$, where $F_{\mathrm{cons/diss}}^{\mu}$ is sourced by the ``regularised'' field $\bigl(h_{\mu\nu}^{\mathrm{ret}}\pm h_{\mu\nu}^{\mathrm{adv}}-h_{\mu\nu}^{\mathrm{S}}\bigr)/2$. What one calls the conservative dynamics is thus described by the equation of motion
\begin{equation}
    m\frac{D^{2}z^{\mu}}{\mathrm{d}\tau^{2}}=F_{\mathrm{cons}}^{\mu}\left(z\right).
\end{equation}
$F_{\mathrm{diss}}^{\mu}$ is responsible for the loss of orbital energy and angular momentum due to back-reaction, which drives the gradual inspiral into the supermassive BH; in contrast, $F_{\mathrm{cons}}^{\mu}$ perturbs the orbital parameters, but does not cause a secular decay of the orbit, being responsible for a slower-than-usual apsidal advance (secular effects on the phase evolution of the system).

As we mentioned, in the framework of binary inspiral and merger different approximate methods necessarily come into play -- usually patched together into a single description by the effective one-body formalism. Conservative GSF effects, in particular, provide a useful strong-field benchmark to make comparison with other methods or to inform higher-order terms that are difficult to determine. In the following, we propose to use predictions of GSF theory in a more broad sense, namely to compare and test theories of gravity. But, before doing so, we pause for a moment the main discussion in order to motivate and introduce ETGs.

\section{Extended Theories of Gravity\label{sec:ETGs}}

Gravity is allegedly the most poorly understood among fundamental interactions. Seeking desperately the guidance of experimental facts, GWs turn research into a data-driven field, letting facts take the lead in the development of new concepts and potentially providing invaluable insights into fundamental physics. Despite the fact that groundbreaking predictions of GR -- such as the light deflection, the Shapiro time delay, the precession of perihelia and the Nordtvedt effect -- passed experimental tests with flying colours, it remains a classical theory which is not expected to be a fundamental one. Indeed, GR has received little direct experimental verification in the strong-field regime, leaving room for alternative or extended theories which reduce to GR in a weak-field limit. Hence, scientists' concern is naturally about seeking a quantum description of spacetime and gravity.

Thus proposing a tentative semi-classical description of gravity may turn out to be a crucial attempt in order to shed light on the full problem of quantum gravity. In this direction, one of the most fruitful approaches comes from ETGs, a broad class of theories which enlarge Einstein's theory by adding higher-order curvature invariants, drawing their motivation from effective quantum-gravity actions \cite{Birrell1982}, and/or scalar fields that can be non-minimally coupled to gravity in the action. Desirable advantages coming from this approach reside in the the fact that ETGs can: (\textit{i}) act as effective field theories for describing certain effects; (\textit{ii}) provide a framework for obtaining predictions for binary evolution and waveforms; and (\textit{iii}) combine constraints coming from the strong-gravity regime with bounds from, say, weak-field regime, cosmology, astrophysics and laboratory tests, separating out the UV from the IR physics. Among the copious possibilities, one of the simplest ways to modify GR is to introduce a scalar field that is non-minimally coupled to gravity, leading to a scalar-tensor (ST) theory of gravity. ST gravity benefits from a well-posed Cauchy formulation for the metric \cite{Salgado2008} -- a necessary feature in order to be amenable for numerical simulations -- as higher-than-second order derivatives of the metric do not come into play. Furthermore, the special class of ST theories with $b=0$ \footnote{$b$ is more typically written as $\omega$ in literature.} in Eq. \eqref{eq:ST_action_JF} is equivalent to metric $f\left(R\right)$ theories \cite{Chiba2003, Wands1993}, which replace the standard GR Lagrangian $\sqrt{-g}R$ with $\sqrt{-g}f\left(R\right)$, $f\left(R\right)$ being a generic function (provided it is analytic, as generally assumed) of the Ricci scalar.

\subsection{Self-force in Scalar-Tensor gravity}
Following \cite{Zimmerman2015}, the action for a generic ST theory in the Jordan frame can be written in full generality as
\begin{align}
    \mathcal{S}= & \frac{1}{16\pi}\int\mathrm{d}^{4}x\sqrt{-\bar{g}}\left[a\left(\bar{\phi}\right)\bar{R}-b\left(\bar{\phi}\right)\bar{g}^{\mu\nu}\bar{\nabla}_{\mu}\bar{\phi}\bar{\nabla}_{\nu}\bar{\phi}-2c\left(\bar{\phi}\right)\right] \nonumber \\
     & +\mathcal{S}_{\mathrm{M}}\left(\bar{g}^{\mu\nu},\Psi\right), \label{eq:ST_action_JF}
\end{align}
where $\Psi$ collectively denotes matter fields, while $a$, $b$ and $c$ are field-dependent ST parameters and the over-bar indicates that we are working within the Jordan frame. Nothing prevents us from redefining the scalar fields in such a way that $a\left(\bar{\phi}\right)\mapsto\bar{\phi}$, leaving the theory with two free functions: the coupling $b$ and the cosmological function $c$. The coupling is responsible for the scalarization phenomenon \cite{Damour1993}, while the cosmological function provides the scalar field with mass and plays the role of the cosmological constant $\Lambda$. In ST gravity the properties of the small body are generally influenced by the presence of the scalar field, due to the variability of the Newton's gravitational constant -- which we will discuss later. Choosing a point-particle action,
\begin{equation}
    \mathcal{S}_{\mathrm{M}}^{\mathrm{pp}}=-\int_{\gamma}m\left(\bar{\phi}\right)\mathrm{d}\bar{\tau},
\end{equation}
we can incorporate this dependence by allowing the mass to vary with $\phi$. Moving to the Einstein frame via a conformal transformation \cite{Damour1996}, the action is cast in the form
\begin{align}
    \mathcal{S}= & \frac{1}{16\pi}\int\mathrm{d}^{4}x\sqrt{-g}\left[R-g^{\mu\nu}\nabla_{\mu}\chi\nabla_{\nu}\chi-2F\left(\chi\right)\right] \nonumber \\
     & -\int_{\gamma}A\left(\chi\right)m\left(\chi\right)\mathrm{d}\tau.    \label{eq:ST_action_EF}
\end{align}
A variation with respect to the metric field yields the field equations
\begin{equation}
    G_{\mu\nu}=8\pi\left(T_{\mu\nu}^{\mathrm{bulk}}+T_{\mu\nu}^{\mathrm{pp}}\right),
\end{equation}
where $T_{\mu\nu}^{\mathrm{bulk}}$ is the stress-energy tensor associated with the bulk scalar field and $T_{\mu\nu}^{\mathrm{pp}}$ is the stress-energy tensor associated with the point particle. Varying the action with respect to the scalar field produces the scalar wave equation
\begin{equation}
    \square\chi-F'\left(\chi\right)=8\pi\int_{\gamma}\frac{\mathrm{d}Am}{\mathrm{d}\chi}\delta_{4}\left(x,z\right)\mathrm{d}\tau, \label{Klein-Gordon-like}
\end{equation}
governing the evolution of the scalar field; here $\square\equiv g^{\mu\nu}\nabla_{\mu}\nabla_{\nu}$ and a prime denotes a derivative with respect to $\chi$; $\delta_{4}\left(x,z\right)\equiv\delta^{4}\left(x^{\mu}-z^{\mu}\right)/\sqrt{-g}$ is the invariant four-dimensional delta-function. One can consider the perturbation of the fields around their background values caused by the point particle, \textit{i.e.} $g_{\mu\nu}=g_{\mu\nu}^{0}+h_{\mu\nu}$ and $\chi=\Phi+f$ -- with $g_{\mu\nu}^{0}\equiv g_{\mu\nu}\left(0\right)$ and $\Phi\equiv\chi\left(0\right)$ denoting the background fields (taken at $m=0$) -- and then write the linearised field equations and the equations of motion.

In light of the general results obtained by Sotiriou and Faraoni \cite{Sotiriou2012},  stationary and axisymmetric BHs are solution of the class of ST theories if and only if they are solutions of GR. The condition that the BH be isolated can be translated into the asymptotic-flatness requirement. In the Jordan frame, this yields $c\left(\bar{\phi}_{0}\right)=0$, where $\bar{\phi}_{0}$ denotes a constant configuration for the scalar field, and $\left.c'\left(\bar{\phi}\right)\right|_{\bar{\phi}_{0}}=0$. In the Einstein frame, these conditions read $\Phi=\Phi_{0}$ and $F\left(\Phi_{0}\right)=\left.F'\left(\Phi\right)\right|_{\Phi_{0}}=0$, ensuring that, for a stationary BH spacetime with a constant scalar field, the next-to-leading order motion is determined by the first-order GSF, which is given by \cite{Zimmerman2015}
\begin{align}
    F_{\mathrm{self}}^{\mu}= & \frac{1}{2}\mathsf{m}w_{\beta}^{\alpha}\left(\nabla^{\beta}h_{\gamma\delta}^{\mathrm{tail}}-2\nabla_{\gamma}h_{\delta}^{\mathrm{tail},\beta}\right)u^{\gamma}u^{\delta} \nonumber \\
     & +\mathsf{q}w_{\beta}^{\alpha}\nabla^{\beta}f_{\mathrm{tail}}. \label{SF in ETGs}
\end{align}
The asymptotic flatness implies the absence of coupling terms between geometry and scalar perturbations, greatly simplifying the equations.

At the end of the day, the result to take home is that the equation of motion \eqref{SF in ETGs} decomposes into a sum of tail contributions,
\begin{equation}
    \mathsf{m}a^{\mu}=F_{\mathrm{tail}}^{\mu}=F_{\mathrm{g,tail}}^{\mu}+F_{\mathrm{s,tail}}^{\mu},
\end{equation}
given by the gravitational and scalar self-force, both depending on the past history of the particle. Here, $\mathsf{m}\equiv m\left(\Phi\right)A\left(\Phi\right)$ is the ST mass parameter.

\subsection{Effective gravitational constant\label{subs:Eff_grav_const}}
An important consequence which stems from the Klein-Gordon-like equation \eqref{Klein-Gordon-like} is that the standard Newtonian potential gets modified by Yukawa-like corrections: the additional degrees of freedom in the gravitational action introduce an effective mass term, introducing terms which smear out the $1/r$ behaviour and disappear at spatial infinity, therefore allowing to recover the Newtonian limit and the cherished Minkowski spacetime. This occurrence was first noticed by Stelle \cite{Stelle1978}, who added terms proportional to $R_{\mu\nu}R^{\mu\nu}$ and $R^{2}$ to the Lagrangian; a recent review with an eye on astronomy and cosmology can be found in \cite{Capozziello2012}.

Quite in general, a great deal of ETGs admit a weak-field limit that, momentarily restoring SI units, can be expressed as
\begin{equation}
    \Phi\left(r\right)=-\frac{GM}{r}\left(1+\sum_{k=1}^{n}\alpha_{k}e^{-r/r_{k}}\right)\equiv-\frac{G_{\mathrm{eff}}M}{r}
\end{equation}
where $r_{k}$ is a characteristic length-scale for the interaction introduced by the $k$-th component of the non-Newtonian corrections, and $G_{\mathrm{eff}}$ is the effective gravitational constant. The amplitude $\alpha_{k}$ of each component is appropriately normalised to the standard Newtonian term and its sign indicates whether the correction is attractive or repulsive. As the Yukawa-like terms are negligible for $r\gg\max r_{k}$, $G$ stands for the gravitational constant as measured at the spatial infinity. Also note that in this picture the inverse-square law holds, as the changes with respect to GR are encoded in the coupling $G_{\mathrm{eff}}$. In general, any correction will introduce a characteristic range, acting at a certain length-scale, which can be translated into the mass $m_{k}$ of a pseudo-particle, whose Compton's length is given by $r_{k}=\hbar/m_{k}c$. From a quantum-field point of view, this fact suggests that, in the low-energy limit, effective theories attempting to unify gravity with other interactions introduce massive particles which carry the gravitational interaction along with the massless graviton.
Taking into account just one component -- or, alternatively, the leading-order term of the summation -- we have
\begin{equation}
    \Phi\left(r\right)=-\frac{GM}{r}\left(1+\alpha e^{-r/\lambda}\right).
\end{equation}
In a nutshell, the effect of a non-Newtonian term can be conveniently parametrised by $\left(\alpha,\lambda\right)$. For distances $r\gg\lambda$, the gravitational coupling is $G$; when $r\ll\lambda$, one can linearise the last equation and get $\Phi\left(r\right)\rightarrow -GM\left(1+\alpha\right)/r$, which, compared to the gravitational force measured in the laboratory, implies $G_{\mathrm{lab}}\simeq G\left(1+\alpha\right)$. Here $G_{\mathrm{lab}}=6.674\ 30\left(15\right)\times10^{-11}\ \mathrm{m^{3}kg^{-1}s^{-2}}$ denotes the usual Newton's constant as measured by Cavendish-like experiments. Of course, $G\equiv G_{\mathrm{lab}}$ in standard GR gravity.

\begin{figure}
    \centering
    \includegraphics[width=0.46\textwidth]{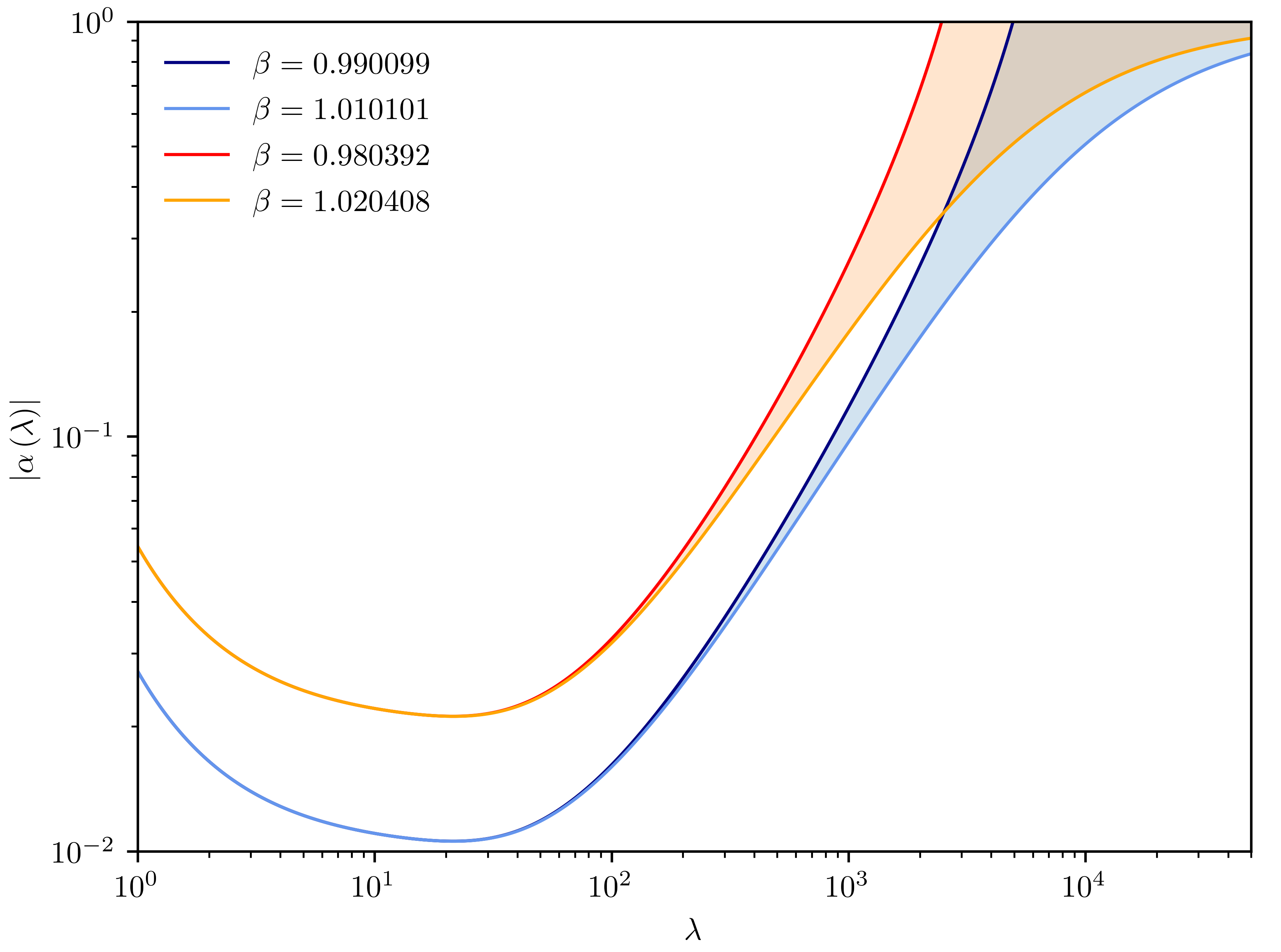}
    \caption{$\left|\alpha\right|$ plotted against $\lambda$ in a log-log scale for Eq. \eqref{eq:alpha(lambda)}. We plotted $\left|\alpha\right|$ against $\lambda$ in order to have positive values only and adopt a log-log scale.}
    \label{fig:alpha}
\end{figure}

The variability of the effective gravitational constant, as given by $G_{\mathrm{eff}}\left(r\right)\equiv G\left(1+\alpha e^{-r/\lambda}\right)$, implies that a measured value of the ratio $G\left(r_{1}\right)/G\left(r_{2}\right)\equiv\beta$ at some distances $r_{1}$ and $r_{2}$ constrains $\alpha$ and $\lambda$ to lie on a curve in the $\left(\alpha,\lambda\right)$ plane, \textit{i.e.}
\begin{equation}
    \alpha\left(\lambda\right)=\frac{\beta-1}{e^{-r_{2}/\lambda}-\beta e^{-r_{1}/\lambda}}. \label{eq:alpha(lambda)}
\end{equation}
Varying $\beta$ within the limits of experimental uncertainties causes the curve to sweep out an allowed region. In Fig. \ref{fig:alpha} we have used four values of $\beta$, corresponding to variations of $\pm1\%$ and $\pm2\%$ of $G_{\mathrm{lab}}$. In this plot, $\alpha$ is expected to lie between the curves, in the shaded area.
Note that, due to the singularity in Eq. \eqref{eq:alpha(lambda)}, neither curve can be extended indefinitely for small and large $\lambda$ -- of course, one should be comfortable with that, as these are only effective parameters.

In conclusion, there is room for experimentally constraining $\lambda$ and $\alpha$, and this has been actually done in the range $1\ \mathrm{cm}<r<10^{8}\ \mathrm{cm}$ with a good deal of techniques \cite{Speake1988, Romaides1994}, giving the estimates $\left|\alpha\right|\lesssim10^{-2}$ and $\lambda\gtrsim10^{2}\ \mathrm{m}$ for the parameters. Binary-pulsar experiments \cite{Taylor1993, Stairs1998, Damour1991} provided a limit in the range $10^{-2}\lesssim\left|\alpha\right|\lesssim10^{-4}$.

\section{Conservative effects\label{sec:ConsEff}}

We now set the stage for analysing the shift of the innermost stable circular orbit (ISCO) in a Schwarzschild background and addressing the GSF-induced modification of the spin-precession rate. Then, we build on our previous discussion to consider these effects within the framework of ETGs. Conservative effects may happen to be buried underneath dissipative effects; however, this by no means implies that our analysis becomes less useful, since the conservative part of the GSF does influence the positional elements of the orbit, \textit{i.e.} parameters containing phase information of the orbit and determining physical attributes such as the (time-dependent) direction of the periapsis and orientation of the orbital plane \cite{Barack2010}.

Consider a bound orbit of a non-spinning, point-like particle with mass $m$ around a Schwarzschild BH of mass $M\gg m$. In the limit $m\rightarrow0$ the trajectory is a timelike geodesic of the background spacetime. The four Killing vectors of the metric lead to the conservation of the orbital energy $E\equiv-u_{t}$, and the direction and magnitude of the angular momentum $L\equiv u_{\phi}$ of the particle, $u^{\mu}=\mathrm{d}z^{\mu}/\mathrm{d}\tau$ being its four-velocity. The conserved quantities $E$, $L$ and $g_{\mu\nu}u^\mu u^\nu =-1$ provide a convenient way to understand the orbits; indeed, one can write
\begin{align}
    \frac{\mathrm{d}t}{\mathrm{d}\tau} = & E\left(1-\frac{2M}{r}\right)^{-1}, \\
    \frac{\mathrm{d}\phi}{\mathrm{d}\tau} = & \frac{L}{r^{2}}, \\
    \left(\frac{\mathrm{d}r}{\mathrm{d}\tau}\right)^{2} = & E^{2}-V\left(r,L\right), \label{eq:Schw.rad.eq.} \\
    \mathrm{with}\hspace{1em}V\left(r,L\right)\equiv & \left(1-\frac{2M}{r}\right)\left(1+\frac{L^{2}}{r^{2}}\right). \label{eq:V(r,L)}
\end{align}
The motion in the Schwarzschild spacetime is governed by the 1D radial equation in the effective potential $V$; Eq. \eqref{eq:Schw.rad.eq.} can be differentiated with respect to the proper time $\tau$ and put into the form
\begin{equation}
    \frac{\mathrm{d}^{2}r}{\mathrm{d}\tau^{2}} = -\frac{1}{2}\frac{\partial V\left(r,L\right)}{\partial r}. \label{eq:Schw.rad.eq.IIorder}
\end{equation}
Before proceeding with the discussion, we mention that the orbital evolution of an EMRI can be qualitatively divided into \cite{Ori2000}:
\begin{itemize}
    \item the \textit{adiabatic inspiral} regime, during which the small body gradually descends through a sequence of geodesic orbits with slowly-varying ``constants'' of motion due to the radiation reaction, which carries orbital energy and angular momentum away from the body;
    \item the \textit{transition} regime, during which the character of the orbit gradually changes as the body nears the ISCO;
    \item the \textit{plunge} regime, during which radiation reaction becomes less important and the small body plunges into the horizon of the central BH.
\end{itemize}
Throughout the late inspiral and transition regimes, the body moves along a nearly-circular orbit, which guarantees that radiated energy and angular momentum are related by $$E=E_{\mathrm{isco}}+\Omega\xi,\hspace{0.6cm}\xi\equiv L-L_{\mathrm{isco}},$$ where $\Omega$ is the orbital angular velocity of the body and $L_{\mathrm{isco}}\equiv\sqrt{12}M$ is the angular momentum at the ISCO. This means that we can regard the effective potential \eqref{eq:V(r,L)} as a function of $r$ and $\xi$, as depicted in Fig. \ref{fig:V(r,xi)}.
\begin{figure}
    \centering
    \includegraphics[width=0.46\textwidth]{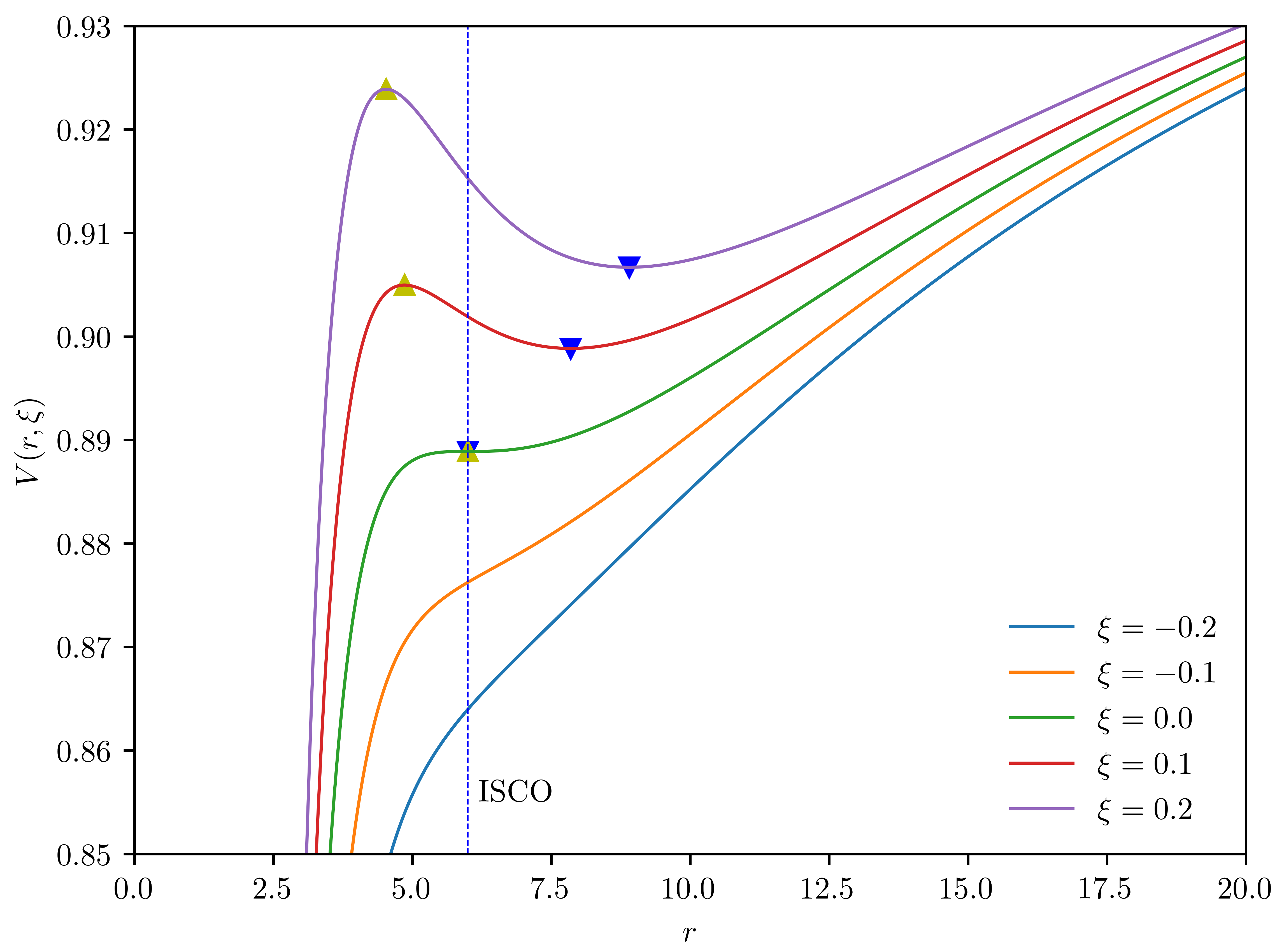}
    \caption{The gradually changing effective potential for radial geodesic motion, plotted for different values of $\xi\equiv L-L_{\mathrm{isco}}$. As $\xi$ decreases to zero, the body sits at the minimum of the potential (denoted by $\blacktriangledown$) and the maximum ($\blacktriangle$) migrates toward larger radii; as $\xi$ nears zero (ISCO), extrema merge; finally, as $\xi$ becomes negative, no extrema occur and the body plunges toward the central BH with nearly constant energy and angular momentum.}
    \label{fig:V(r,xi)}
\end{figure}

\subsection{ISCO shift}
When dealing with a slightly eccentric orbit that represents an $e$-perturbation of a circular orbit with radius $r_{\circ}$, one can write $r\left(\tau\right)=r_{\circ}+er_{1}\left(\tau\right)+O\left(e^{2}\right)$, assuming that $r_{1}\left(\tau\right)$ is $e$-independent. Substituting in Eq. \eqref{eq:Schw.rad.eq.IIorder} and performing linear variations with respect to $e$ (holding $r_\circ$ fixed), one can show \cite{BarackSago2010} that the $O\left(e\right)$ radial motion is simple-harmonic in $\tau$ with frequency
\begin{equation}
    \omega_{r}^{2}\equiv\frac{1}{2}\left.\frac{\partial^{2}V\left(r,L\right)}{\partial r^{2}}\right|_{e=0}=\frac{M\left(r_{\circ}-6M\right)}{r_{\circ}^{3}\left(r_{\circ}-3M\right)}.
\end{equation}
The orbit is stable under small-$e$ perturbations when $\omega_{r}^{2}>0$, namely for $r_{\circ}>6M$, and is perturbatively unstable when $\omega_{r}^{2}<0$. The ISCO is identified by the defining condition $\omega_{r}^{2}=0$, giving $r_{\mathrm{isco}}=6M$.

Dealing with the finiteness of $m$ (yet still much smaller than $M$), $E$ and $L$ are not expected to be constant of the motion any longer; indeed, the equations of motion become \cite{BarackSago2010}
\begin{equation}
    \frac{\mathrm{d}\tilde{E}}{\mathrm{d}\tilde{\tau}}=-\frac{F_{t}^{\mathrm{cons}}}{m},\hspace{1em} \frac{\mathrm{d}\tilde{L}}{\mathrm{d}\tilde{\tau}}=\frac{F_{\phi}^{\mathrm{cons}}}{m},
\end{equation}
\begin{equation}
    \frac{\mathrm{d}^{2}\tilde{r}}{\mathrm{d}\tilde{\tau}^{2}}=-\frac{1}{2}\frac{\partial V\left(\tilde{r},\tilde{L}\right)}{\partial \tilde{r}}+\frac{F_{\mathrm{cons}}^{r}}{m}, \label{eq:radial_eq.+GSF}
\end{equation}
where tilded quantities refer to the GSF-corrected orbit (\textit{i.e.} in the perturbed spacetime $\tilde{g}_{\mu\nu}$, Eq. \eqref{eq:tildeg}). Specialising to a slightly eccentric and GSF-perturbed orbit, similarly to what we did before we write $\tilde{r}\left(\tilde{\tau}\right)=r_{\circ}+e\tilde{r}_{1}\left(\tilde{\tau}\right)+O\left(e^{2}\right)$. Then, through $O\left(e\right)$, Eq. \eqref{eq:radial_eq.+GSF} entails a simple-harmonic motion with frequency
\begin{equation}
    \tilde{\omega}_{r}^{2}\equiv\left.\frac{\mathrm{d}}{\mathrm{d}\tilde{r}}\left[\frac{1}{2}\frac{\partial V\left(\tilde{r},\tilde{L}\right)}{\partial \tilde{r}}-\frac{F_{\mathrm{cons}}^{r}\left(\tilde{r}\right)}{m}\right]\right|_{\tilde{r}=r_{\circ}}.
\end{equation}
An explicit expression for the GSF-shifted radial frequency $\tilde{\omega}_{r}$, describing the $O\left(m\right)$ conservative shift in the radial frequency off its geodesic value, has been worked out by \cite{BarackSago2010}.

The defining condition for the ISCO now reads $\tilde{\omega}_{r}^{2}\left(r_{\circ}=\tilde{r}_{\mathrm{ISCO}}\right)=0$ through $O\left(m\right)$. The GSF shifts the ISCO both in frequency and location. We will not consider the shift in location as, under a gauge displacement $x^{\mu}\longrightarrow x'^{\mu}=x^{\mu}+\xi^{\mu}\left(x\right)$, the radial coordinate -- being simply a coordinate -- is not left invariant and it is not useful as a benchmark. We rather consider the (GSF-corrected) circular-orbit azimuthal frequency,
\begin{equation}
    \tilde{\Omega}\equiv\frac{\mathrm{d}\tilde{\phi}}{\mathrm{d}\tilde{t}}=\left(\frac{\mathrm{d}\tilde{t}}{\mathrm{d}\tilde{\tau}}\right)^{-1}\frac{\mathrm{d}\tilde{\phi}}{\mathrm{d}\tilde{\tau}}=\frac{\tilde{u}^{\phi}}{\tilde{u}^{t}},
\end{equation}
which can be seen to be invariant under all $O\left(m\right)$ gauge transformations whose generators respect the helical symmetry of the circular-orbit configuration \cite{Sago2008}. It has been shown that the GSF-induced frequency shift $\Delta\Omega\equiv\tilde{\Omega}-\Omega$ through $O\left(m\right)$ reads \cite{BarackSago2010}
\begin{align}
    \Delta\Omega_{\mathrm{isco}} & \equiv \tilde{\Omega}\left(\tilde{r}_{\mathrm{isco}}\right)-\Omega\left(r_{\mathrm{isco}}\right) \label{eq:Delta_Omega}\\
    & = -\frac{1}{6^{3/2}M}\left[\frac{\Delta r_{\mathrm{isco}}}{4M}+\frac{27M}{2m}F_{\circ\mathrm{isco}}^{r}\right], \label{eq:Delta_Omega_2}
\end{align}
where $\Delta r_{\mathrm{isco}}\equiv\tilde{r}_{\mathrm{isco}}-6M$, $F_{\circ\mathrm{isco}}^{r}\equiv F_{\circ}^{r}\left(r_{\circ}=6M\right)$ is the circular-orbit value of the conservative GSF evaluated at the ISCO, and $\left(6^{3/2}M\right)^{-1}=\Omega_{\mathrm{isco}}$  is the geodesic value of the orbital frequency at $r=6M$. Actually, despite its gauge-invariant character, as pointed out in \cite{Barack2005} $\Delta\Omega_{\mathrm{isco}}$ suffers from a pathology of the Lorenz-gauge metric perturbation which is not asymptotically flat. Hence one uses a ``rescaled'' frequency, given through $O\left(m\right)$ by $\hat{\tilde{\Omega}}=\left(1-\alpha\right)\tilde{\Omega}$. This modifies Eq. \eqref{eq:Delta_Omega_2} to
\begin{equation}
    \Delta\hat{\Omega}_{\mathrm{isco}}=-\frac{1}{6^{3/2}M}\left[\frac{\Delta r_{\mathrm{isco}}}{4M}+\frac{27}{2}\frac{M}{m}F_{\circ\mathrm{isco}}^{r}+\frac{1}{\sqrt{18}}\frac{m}{M}\right].
\end{equation}
This orbital-frequency shift has been numerically evaluated, yielding \cite{BarackSago2010}
\begin{equation}
    \Delta\hat{\Omega}_{\mathrm{isco}}=\Omega_{\mathrm{isco}}\times0.251\ 2\left(4\right)\eta,    \label{eq:DeltaOmega_numerical}
\end{equation}
where, recall, $\eta\equiv m/M$ is the mass ratio of the system.

\subsection{Spin precession and self-torque}
The geodesic precession is a familiar effect of GR dynamics, associated with the failure of a spin vector to return to itself after being parallel-transported along a closed curve in a curved spacetime. This effect was first quantified in the context of GSF by \cite{Dolan2014}, who considered a slowly spinning body of mass $m$ in a circular orbit around a Schwarzschild BH of mass $M\gg m$ and calculated the $O\left(\eta\right)$ shift in the precession rate due to the back-reaction of the conservative piece of the GSF. Note that the geodesic value of the spin precession rate is found to be $$\psi=1-\sqrt{1-\frac{3M}{r_{\Omega}}},$$ so that it can be as large as $\sim 0.3$ at the ISCO: this is some 100$^{\circ}$ rotation of the spin axis over a single orbital period.

In the test-particle limit ($m\rightarrow0$ and back-reaction is negligible), assuming that the spin $s_{\mu}$ is non-zero, but sufficiently small so as not to affect the motion, one can state that the particle follows a timelike geodesics of the unperturbed metric $g_{\mu\nu}$, \textit{i.e.}
\begin{equation}
    u^{\mu}\nabla_{\mu}u^{\nu}=0 \hspace{1em} \mathrm{and} \hspace{1em} u^{\mu}\nabla_{\mu}s^{\nu}=0,
    \label{eq:parallel_transport in g}
\end{equation}
where $u^\mu$ is the four-velocity of the particle and $\nabla_\mu$ is the covariant derivative compatible with $g_{\mu\nu}$. The precession effect is encoded in Eq. \eqref{eq:parallel_transport in g}.

The geodesic effect gets perturbed when one endows the particle with a mass $m$ (yet still much smaller than $M$). As discussed in Sec. \ref{sec:GSF}, Eqs. \eqref{eq:parallel_transport in g} remain valid through $O\left(m\right)$ if one replaces the background metric with the effective metric $\tilde{g}_{\mu\nu}=g_{\mu\nu}+h_{\mu\nu}^{\mathrm{R}}$, where one can write
\begin{equation}
    \tilde{u}^{\mu}\tilde{\nabla}_{\mu}\tilde{u}^{\nu}=0 \hspace{1em} \mathrm{and} \hspace{1em} \tilde{u}^{\mu}\tilde{\nabla}_{\mu}\tilde{s}^{\nu}=0,
\end{equation}
where the tilde denotes that we are now referring to quantities evaluated with respect to $\tilde{g}_{\mu\nu}$. In the spirit of GR equivalence principle, we can say that, just as geodesic motion in $\tilde{g}_{\mu\nu}$ corresponds to self-acceleration in $g_{\mu\nu}$, parallel transport of $\tilde{s}_{\mu}$ in $\tilde{g}_{\mu\nu}$ corresponds to self-torque in $g_{\mu\nu}$. The GSF correction has been encoded in $\delta\psi\equiv\tilde{\psi}-\psi$ by \cite{Dolan2014}, which is a function of the invariant quantity $\Omega$ and can be computed directly from the knowledge of $\nabla_{\mu}h_{\nu\rho}^{\mathrm{R}}$ on the worldline.

\subsection{ISCO shift and spin precession in ETGs}
The asymptotic flatness of the Schwarzschild spacetime guarantees that the GSF in a broad class of ETGs -- ST and (a subset of) $f\left(R\right)$ theories -- takes the form \eqref{SF in ETGs}; this has to be compared to Eq. \eqref{eq:MiSaTaQuWa_R} in GR.

Some remarks are necessary at this point. \textit{First}, the tail field does not satisfy any particular field equation, nor is it differentiable on the worldline \cite{PoiPouVeg2011}. As we mentioned, Detweiler and Whiting provided a more compelling form of the MiSaTaQuWa equation thanks to a S+R decomposition of the retarded field, whose advantage is in bolstering the interpretation as \textit{self}-fields. But it should be understood that neither of these two fields represents the actual physical perturbation from the particle, which is of course the retarded field; hence both descriptions of the perturbed motion are equally valid interpretation of the same physical effect and there is nothing wrong with adopting one or the other decomposition -- both produce the same final numerical value for the GSF.\footnote{The only aspect demanding attention is that statements referring to the smoothness of the R-field would need to be formulated more carefully to reflect the irregularity in higher derivatives of the tail field, \cite{Barack2009}.}
\textit{Second}, Eq. \eqref{SF in ETGs} resides in the Einstein frame, while Eq. \eqref{eq:MiSaTaQuWa_R} is in the Jordan frame. This might seem a serious, physical concern, suggesting that the Einstein-frame action \eqref{eq:ST_action_EF} may not describe the same physics as the Jordan-frame action \eqref{eq:ST_action_JF} does. However, it should be clear that those are just different \textit{representations} of the \textit{same} theory \cite{Sotiriou2008}, and whether or not we choose to represent a theory with respect to a certain metric is, after all, simply irrelevant.

Coming back to the point of our discussion, in ETGs, even with a constant background scalar field, the scalar perturbation is also expected to influence the motion of a point particle orbiting a ST-BH, since it would experience both metric and scalar tail pieces of the GSF. At the end of the day, these deviations from GR would likely have observational consequences. In order to establish a full comparison between competing theories, one can imagine to set up a numerical calculation of the GSF on a point particle around a BH in ST gravity; but doing so is, in general, a tremendously complicated task, well beyond the scope of this paper. Hence, we will rather follow the path outlined in subsection \ref{subs:Eff_grav_const}. In order to account for the variability of Newton's constant, we restore $G$ and $c$ in the equations; then we replace $G\longmapsto G\left(1+\alpha\right)$ and, finally, set $G=c=1$ again.

So our estimate of the frequency at the ISCO reads $$\Omega_{\mathrm{isco}}^{\mathrm{ETG}}=\frac{1}{6^{3/2}M\left(1+\alpha\right)}=\frac{\Omega_{\mathrm{isco}}^{\mathrm{GR}}}{1+\alpha}.$$ Our \textit{ansatz} is that the shift in the orbital frequency retains its functional form in passing from GR to ETGs, \textit{i.e.} $\Delta\Omega_{\mathrm{isco}}^{\mathrm{ETG}}\equiv\Delta\Omega_{\mathrm{isco}}^{\mathrm{GR}}\left(G_{\mathrm{eff}}\right)$; recalling Eq. \eqref{eq:DeltaOmega_numerical}, this yields
\begin{equation}
    \frac{\Delta\Omega_{\mathrm{isco}}^{\mathrm{ETG}}}{\Omega_{\mathrm{isco}}^{\mathrm{GR}}}=\frac{0.251\ 2\left(4\right)\eta}{1+\alpha}.
\end{equation}

\begin{figure}
    \centering
    \includegraphics[width=0.46\textwidth]{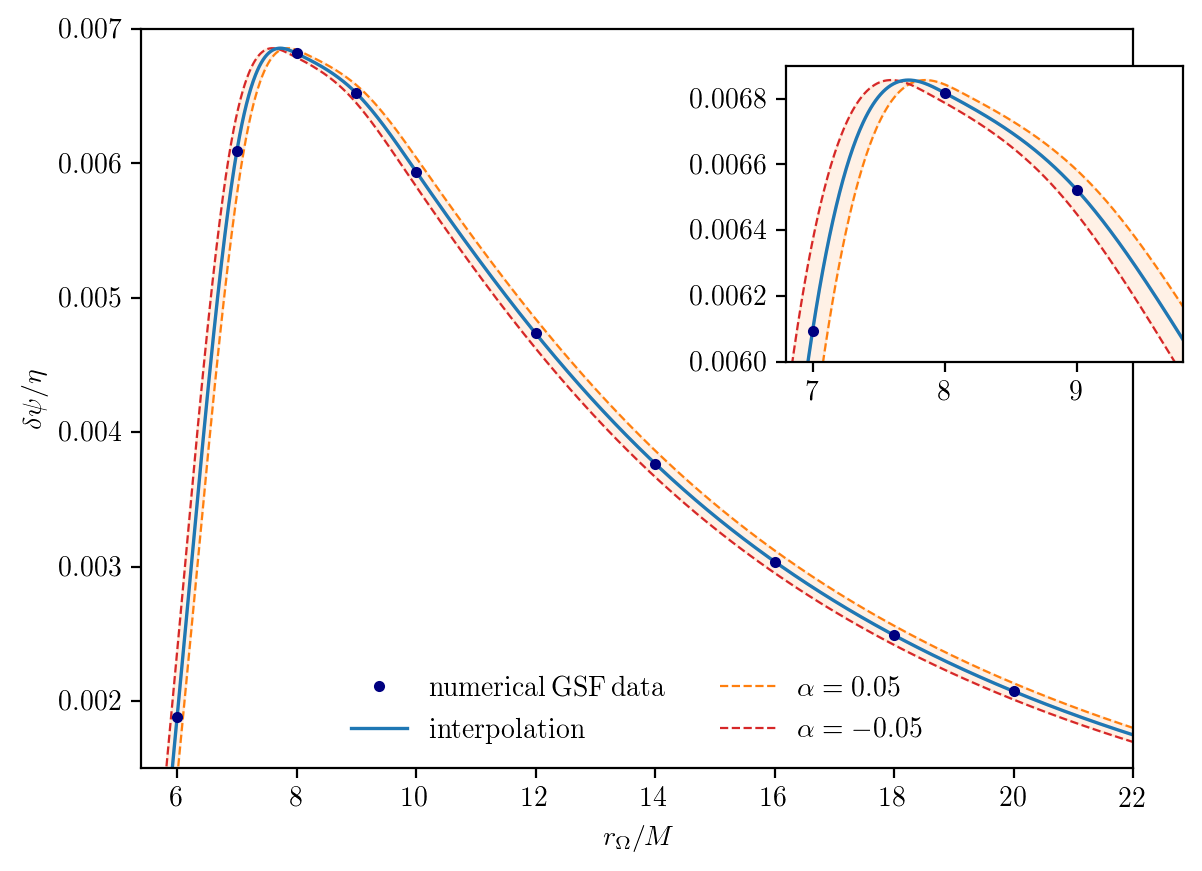}
    \caption{The conservative correction to $\psi$ as a function of the invariant orbital radius $r_{\Omega}$. The solid line interpolates the numerical GSF data \cite{Dolan2014}; the dashed lines show the estimate of deviations from GR for $\alpha=\pm 5\times10^{-2}$; the inset zooms on the peak.}
    \label{fig:spin_precession}
\end{figure}

In a similar fashion we can deal with the spin-precession shift. Following \cite{Dolan2014}, we plot $\delta\psi/\eta$ against $r_{\Omega}/M=\sqrt[3]{G/\left(\Omega^{2}M^{2}\right)}$ (in physical units). Repeating the previous reasoning, we get
\begin{equation}
    r_{\Omega}^{\mathrm{ETG}}=r_{\Omega}^{\mathrm{GR}}\sqrt[3]{1+\alpha}    \label{eq:invariant_radius_ETGs}
\end{equation}
for the gauge-invariant radius. In Fig. \ref{fig:spin_precession} we plot numerical values for $\delta\psi$ and compare with the shifted result given by Eq. \eqref{eq:invariant_radius_ETGs}; the solid line is the interpolating cubic spline for the GSF data, intended only as a guide to the eye since no model is assumed.

\section{Detectability\label{sec:S/N}}
Finally, let us move on to address detectability of EMRIs. In general, the number of cycles generated at the frequency $f$ can be estimated as $$N_{\mathrm{cycles}}\left(f\right)=\frac{f}{2\pi}\frac{\mathrm{d}\phi}{\mathrm{d}f}=\frac{f}{2\pi}\dot{\phi}\dot{f}^{-1}=\frac{f^2}{\dot{f}},$$ where $\phi$ is the orbital phase. For an inspiral process, the number of cycles in the frequency window $\left(f_1, f_2\right)$ is given by
$$N_{\mathrm{cycles}}=\int_{f_1}^{f_2} \frac{f^2}{\dot{f}} \mathrm{d}f.$$
As mentioned in the introduction, the small body spends $10^{4-5}$ cycles while inspiraling into the supermassive BH, yielding years of observability before the final plunge. The characteristic strain $h_c$ -- which includes the effect of integrating the inspiraling signal -- from the source emitting at a frequency $f$ is
$$h_c\left(f\right) \equiv 2f\left|\tilde{h}\left(f\right)\right|\simeq \sqrt{\frac{2f^2}{\dot{f}}}h_{\mathrm{rms}},$$ where $h_{\mathrm{rms}}$ is the (approximately) constant root-mean-square amplitude of the source signal; this form is derived from the Fourier transform in the stationary-phase approximation \cite{Moore2015}. The characteristic strain can be directly related to the proper distance $d$ and the GW-emission power $\dot{E}$ of the source as $$h_c\left(f\right)=\frac{1}{\pi d}\sqrt{\frac{2\dot{E}}{\dot{f}}}.$$
For EMRIs, this is about $10^{-20}-10^{-19}$. A fully coherent search of $10^{4-5}$ cycles, albeit conceptually straightforward, is computationally unfeasible; indeed, a hierarchical matched filtering approach has to be pursued, dividing  data into shorter time segments and performing a coherent search. Only then can the $S/N$ be built up by incoherently adding the power in the segments \cite{Gair2004}. This procedure reduces the $S/N$ by a factor of $N^{-1/4}$ \cite{Maggiore}, $N$ being the number of segments: while a fully coherent search would require a $S/N \gtrsim 15$ \cite{Babak2017}, an incoherent search would probably detect signals with $S/N \gtrsim 20$. The $S/N$ can be estimated by $$\left(\frac{S}{N}\right)^2 = \int_{f_1}^{f_2}\frac{h^2_c\left(f\right)}{h^2_n\left(f\right)}\mathrm{d}\left(\log f\right),$$ where $h^2_n\left(f\right)\equiv f S_n \left(f\right)$ and $S_n \left(f\right)$ is the noise spectral density of the detector \cite{Moore2015,Maggiore}. One of the most limiting factors for detection of EMRIs is represented by the stochastic foreground from unresolved galactic binaries, which act as an effective -- although not stationary -- noise source. This confusion noise will affect detection less as observation time accumulates and more foreground sources get removed \cite{LISA, Robson2018}. To get an idea of the characteristic strain of sources in the frequency range of LISA compared with its sensitivity, reference can be made to Figure 1 in the proposal paper of the Mission \cite{LISA}, where also five simultaneously evolving harmonics of an EMRI at $z=1.2$ are shown.

Looking at the problem of inspiraling compact binaries (not EMRIs here) it is not difficult to realise the need for a very high-order computation in the post-Newtonian (PN) regime. This reflects the fact that experiments hunt for signals which are buried in a noise orders of magnitude larger than the GW signal itself, needing matched filtering for signal extraction \cite{Maggiore}. In order to track the evolution of the signal accurately enough to allow detection, one needs templates able to reproduce the number of cycles with a precision at least $O\left(1\right)$, implying the inclusion of corrections up to $O\left(v^5/c^5\right)$ -- \textit{i.e.}, PN corrections to the phase at least up to 3PN level, better yet to 3.5PN.

In the framework of ETGs, it has been shown that corrections to GR equations and waveforms describing binary BH systems are negligible up to 2.5PN order \cite{Mirshekari2013}. Furthermore, a study to constrain ETGs based on GW150914 and GW151226 events has been conducted up to 3.5PN level \cite{DeLaurentis2016}: since GW frequencies get modulated through the replacement $G\rightarrow G_{\mathrm{eff}}$, the phase of the waveform changes accordingly. Within a single-parameter analysis, the impossibility of ruling out $\left|\alpha\right|<10^{-2}$ has been reported. However, tighter limits should be set with access to a greater number of events -- parameters of the two binary-system events are not currently estimated with high enough precision. What should be stressed, again, is that this kind of constraints on metric theories of gravity are unprecedented in nature, as they concern a strong-field regime.

Recent work has included GSF corrections to the binding energy of binaries in order to improve phase accuracy \cite{Huerta2017}, thus making use of GSF formalism together with PN theory to robustly capture the inspiral dynamics of compact binaries with asymmetric mass-ratios and reproduce the quasi-circular limit. However, for EMRIs the issue becomes more difficult to deal with, even within GR. Further work should be pursued in the future.

\section*{Discussion and conclusions}
We have followed an already firmly established indication of modifications to the equations of motion in ST gravity and we used a phenomenological approach to incorporate these modifications in a simple and direct way. Clearly, the concept of a theory-agnostic parametrisation relies on the fact that it is very difficult, at present, to discern between GR and non-GR spacetimes -- hence, a theory-dependent modelling would prove to be an unnecessary titanic enterprise. This justifies the phenomenological approach that we pursued, being satisfactory as long as one is content to capture some preliminary expectations as to the modifications one may reasonably expect. Indeed, we have identified a class of effects as a promising test-bed for gravitation theories.
These results motivate the scientific interest in the promising field of the GSF technique, not only for its considerable astrophysical impact, but also for the importance of the additional constraints it could provide on the plethora of ETGs that have been developed so far. It cannot be overemphasised how future space-borne experiments will play a crucial role, since they will open for the first time a wide window on the deep gravitational universe.

\bibliography{biblio}

\begin{thebibliography}{46}
\expandafter\ifx\csname natexlab\endcsname\relax\def\natexlab#1{#1}\fi
\expandafter\ifx\csname bibnamefont\endcsname\relax
  \def\bibnamefont#1{#1}\fi
\expandafter\ifx\csname bibfnamefont\endcsname\relax
  \def\bibfnamefont#1{#1}\fi
\expandafter\ifx\csname citenamefont\endcsname\relax
  \def\citenamefont#1{#1}\fi
\expandafter\ifx\csname url\endcsname\relax
  \def\url#1{\texttt{#1}}\fi
\expandafter\ifx\csname urlprefix\endcsname\relax\def\urlprefix{URL }\fi
\providecommand{\bibinfo}[2]{#2}
\providecommand{\eprint}[2][]{\url{#2}}

\bibitem[{\citenamefont{{{LIGO Scientific Collaboration and Virgo
  Collaboration}}}(2016{\natexlab{a}})}]{LIGO-Virgo2016}
\bibinfo{author}{\bibnamefont{{{LIGO Scientific Collaboration and Virgo
  Collaboration}}}}, \bibinfo{journal}{Phys. Rev. Lett.}
  \textbf{\bibinfo{volume}{116}} (\bibinfo{year}{2016}{\natexlab{a}}),
  \eprint{1602.03837}.

\bibitem[{\citenamefont{{{LIGO Scientific Collaboration and Virgo
  Collaboration}}}(2017)}]{LIGO-Virgo2017}
\bibinfo{author}{\bibnamefont{{{LIGO Scientific Collaboration and Virgo
  Collaboration}}}}, \bibinfo{journal}{Class. Quant. Grav.}
  \textbf{\bibinfo{volume}{34}} (\bibinfo{year}{2017}), \eprint{1611.07531}.

\bibitem[{\citenamefont{Cornish and Robson}(2017)}]{LISA_newdesign}
\bibinfo{author}{\bibfnamefont{N.}~\bibnamefont{Cornish}} \bibnamefont{and}
  \bibinfo{author}{\bibfnamefont{T.}~\bibnamefont{Robson}}
  (\bibinfo{year}{2017}), \eprint{1703.09858}.

\bibitem[{\citenamefont{{{LIGO Scientific Collaboration and Virgo
  Collaboration}}}(2016{\natexlab{b}})}]{aLIGO_transient_noise}
\bibinfo{author}{\bibnamefont{{{LIGO Scientific Collaboration and Virgo
  Collaboration}}}} (\bibinfo{year}{2016}{\natexlab{b}}), \eprint{1602.03844}.

\bibitem[{\citenamefont{Martynov and et~al.}(2016)}]{aLIGO_sensitivity}
\bibinfo{author}{\bibfnamefont{D.~V.} \bibnamefont{Martynov}} \bibnamefont{and}
  \bibinfo{author}{\bibnamefont{et~al.}}, \bibinfo{journal}{Phys. Rev. D}
  \textbf{\bibinfo{volume}{93}} (\bibinfo{year}{2016}), \eprint{1604.00439}.

\bibitem[{\citenamefont{{{LISA L3 Mission Proposal}}}(2017)}]{LISA}
\bibinfo{author}{\bibnamefont{{{LISA L3 Mission Proposal}}}}
  (\bibinfo{year}{2017}), \eprint{1702.00786}.

\bibitem[{\citenamefont{Amaro-Seoane et~al.}(2007)\citenamefont{Amaro-Seoane,
  Gair, Freitag, Miller, Mandel, Cutler, and Babak}}]{Amaro-Seoane2007}
\bibinfo{author}{\bibfnamefont{P.}~\bibnamefont{Amaro-Seoane}},
  \bibinfo{author}{\bibfnamefont{J.~R.} \bibnamefont{Gair}},
  \bibinfo{author}{\bibfnamefont{M.}~\bibnamefont{Freitag}},
  \bibinfo{author}{\bibfnamefont{M.~C.} \bibnamefont{Miller}},
  \bibinfo{author}{\bibfnamefont{I.}~\bibnamefont{Mandel}},
  \bibinfo{author}{\bibfnamefont{C.~J.} \bibnamefont{Cutler}},
  \bibnamefont{and} \bibinfo{author}{\bibfnamefont{S.}~\bibnamefont{Babak}},
  \bibinfo{journal}{Class. Quant. Grav.} \textbf{\bibinfo{volume}{24}}
  (\bibinfo{year}{2007}), \eprint{astro-ph/0703495}.

\bibitem[{\citenamefont{Le~Tiec}(2014)}]{LeTiec2014}
\bibinfo{author}{\bibfnamefont{A.}~\bibnamefont{Le~Tiec}},
  \bibinfo{journal}{Int. J. Mod. Phys. D} \textbf{\bibinfo{volume}{23}}
  (\bibinfo{year}{2014}), \eprint{1408.5505}.

\bibitem[{\citenamefont{Mino et~al.}(1997)\citenamefont{Mino, Sasaki, and
  Tanaka}}]{MiSaTa}
\bibinfo{author}{\bibfnamefont{Y.}~\bibnamefont{Mino}},
  \bibinfo{author}{\bibfnamefont{M.}~\bibnamefont{Sasaki}}, \bibnamefont{and}
  \bibinfo{author}{\bibfnamefont{T.}~\bibnamefont{Tanaka}},
  \bibinfo{journal}{Phys. Rev. D} \textbf{\bibinfo{volume}{55}}
  (\bibinfo{year}{1997}), \eprint{9606018}.

\bibitem[{\citenamefont{Quinn and Wald}(1997)}]{QuWa}
\bibinfo{author}{\bibfnamefont{T.~C.} \bibnamefont{Quinn}} \bibnamefont{and}
  \bibinfo{author}{\bibfnamefont{R.~M.} \bibnamefont{Wald}},
  \bibinfo{journal}{Phys. Rev. D} \textbf{\bibinfo{volume}{56}}
  (\bibinfo{year}{1997}), \eprint{9610053}.

\bibitem[{\citenamefont{Dirac}(1938)}]{Dirac1938}
\bibinfo{author}{\bibfnamefont{P.~A.~M.} \bibnamefont{Dirac}},
  \bibinfo{journal}{Proc. R. Soc. London A} \textbf{\bibinfo{volume}{A167}}
  (\bibinfo{year}{1938}).

\bibitem[{\citenamefont{DeWitt and Brehme}(1960)}]{DeWitt1960}
\bibinfo{author}{\bibfnamefont{B.~S.} \bibnamefont{DeWitt}} \bibnamefont{and}
  \bibinfo{author}{\bibfnamefont{R.~W.} \bibnamefont{Brehme}},
  \bibinfo{journal}{Annals of Phys.} \textbf{\bibinfo{volume}{9}}
  (\bibinfo{year}{1960}).

\bibitem[{\citenamefont{Detweiler and Whiting}(2003)}]{DetWhi2003}
\bibinfo{author}{\bibfnamefont{S.}~\bibnamefont{Detweiler}} \bibnamefont{and}
  \bibinfo{author}{\bibfnamefont{B.~F.} \bibnamefont{Whiting}},
  \bibinfo{journal}{Phys. Rev. D} \textbf{\bibinfo{volume}{67}}
  (\bibinfo{year}{2003}), \eprint{0202086}.

\bibitem[{\citenamefont{Barack and Pound}(2018)}]{BarPou2018}
\bibinfo{author}{\bibfnamefont{L.}~\bibnamefont{Barack}} \bibnamefont{and}
  \bibinfo{author}{\bibfnamefont{A.}~\bibnamefont{Pound}}
  (\bibinfo{year}{2018}), \eprint{1805.10385}.

\bibitem[{\citenamefont{Poisson et~al.}(2011)\citenamefont{Poisson, Pound, and
  Vega}}]{PoiPouVeg2011}
\bibinfo{author}{\bibfnamefont{E.}~\bibnamefont{Poisson}},
  \bibinfo{author}{\bibfnamefont{A.}~\bibnamefont{Pound}}, \bibnamefont{and}
  \bibinfo{author}{\bibfnamefont{I.}~\bibnamefont{Vega}}
  (\bibinfo{year}{2011}), \eprint{1102.0529}.

\bibitem[{\citenamefont{Birrell and Davies}(1982)}]{Birrell1982}
\bibinfo{author}{\bibfnamefont{N.~D.} \bibnamefont{Birrell}} \bibnamefont{and}
  \bibinfo{author}{\bibfnamefont{P.~C.~W.} \bibnamefont{Davies}},
  \emph{\bibinfo{title}{Quantum Fields in Curved Space}}
  (\bibinfo{publisher}{Cambridge University Press}, \bibinfo{year}{1982}).

\bibitem[{\citenamefont{Salgado et~al.}(2008)\citenamefont{Salgado,
  Martinez-del Rio, Alcubierre, and N\'{u}\~{n}ez}}]{Salgado2008}
\bibinfo{author}{\bibfnamefont{M.}~\bibnamefont{Salgado}},
  \bibinfo{author}{\bibfnamefont{D.}~\bibnamefont{Martinez-del Rio}},
  \bibinfo{author}{\bibfnamefont{M.}~\bibnamefont{Alcubierre}},
  \bibnamefont{and}
  \bibinfo{author}{\bibfnamefont{D.}~\bibnamefont{N\'{u}\~{n}ez}},
  \bibinfo{journal}{Phys. Rev. D} \textbf{\bibinfo{volume}{77}}
  (\bibinfo{year}{2008}), \eprint{0801.2372}.

\bibitem[{\citenamefont{Chiba}(2003)}]{Chiba2003}
\bibinfo{author}{\bibfnamefont{T.}~\bibnamefont{Chiba}},
  \bibinfo{journal}{Phys. Lett. B} \textbf{\bibinfo{volume}{575}}
  (\bibinfo{year}{2003}).

\bibitem[{\citenamefont{Wands}(1994)}]{Wands1993}
\bibinfo{author}{\bibfnamefont{D.}~\bibnamefont{Wands}},
  \bibinfo{journal}{Class. Quant. Grav.} \textbf{\bibinfo{volume}{11}}
  (\bibinfo{year}{1994}), \eprint{9307034}.

\bibitem[{\citenamefont{Zimmerman}(2015)}]{Zimmerman2015}
\bibinfo{author}{\bibfnamefont{P.}~\bibnamefont{Zimmerman}},
  \bibinfo{journal}{Phys. Rev. D} \textbf{\bibinfo{volume}{92}}
  (\bibinfo{year}{2015}), \eprint{1507.04076}.

\bibitem[{\citenamefont{Damour and Esposito-Far\`{e}se}(1993)}]{Damour1993}
\bibinfo{author}{\bibfnamefont{T.}~\bibnamefont{Damour}} \bibnamefont{and}
  \bibinfo{author}{\bibfnamefont{G.}~\bibnamefont{Esposito-Far\`{e}se}},
  \bibinfo{journal}{Phys. Rev. Lett.} \textbf{\bibinfo{volume}{70}}
  (\bibinfo{year}{1993}).

\bibitem[{\citenamefont{Damour and Esposito-Far\`{e}se}(1996)}]{Damour1996}
\bibinfo{author}{\bibfnamefont{T.}~\bibnamefont{Damour}} \bibnamefont{and}
  \bibinfo{author}{\bibfnamefont{G.}~\bibnamefont{Esposito-Far\`{e}se}},
  \bibinfo{journal}{Phys. Rev. D} \textbf{\bibinfo{volume}{54}}
  (\bibinfo{year}{1996}).

\bibitem[{\citenamefont{Sotiriou and Faraoni}(2012)}]{Sotiriou2012}
\bibinfo{author}{\bibfnamefont{T.~P.} \bibnamefont{Sotiriou}} \bibnamefont{and}
  \bibinfo{author}{\bibfnamefont{V.}~\bibnamefont{Faraoni}},
  \bibinfo{journal}{Phys. Rev. Lett.} \textbf{\bibinfo{volume}{108}}
  (\bibinfo{year}{2012}), \eprint{1109.6324}.

\bibitem[{\citenamefont{Stelle}(1978)}]{Stelle1978}
\bibinfo{author}{\bibfnamefont{K.~S.} \bibnamefont{Stelle}},
  \bibinfo{journal}{Gen. Relat. Gravit.} \textbf{\bibinfo{volume}{9}}
  (\bibinfo{year}{1978}).

\bibitem[{\citenamefont{Capozziello and De~Laurentis}(2012)}]{Capozziello2012}
\bibinfo{author}{\bibfnamefont{S.}~\bibnamefont{Capozziello}} \bibnamefont{and}
  \bibinfo{author}{\bibfnamefont{M.}~\bibnamefont{De~Laurentis}},
  \bibinfo{journal}{Ann. Phys.} \textbf{\bibinfo{volume}{524}}
  (\bibinfo{year}{2012}).

\bibitem[{\citenamefont{Speake and Quinn}(1988)}]{Speake1988}
\bibinfo{author}{\bibfnamefont{C.~C.} \bibnamefont{Speake}} \bibnamefont{and}
  \bibinfo{author}{\bibfnamefont{T.~J.} \bibnamefont{Quinn}},
  \bibinfo{journal}{Phys. Rev. Lett.} \textbf{\bibinfo{volume}{61}}
  (\bibinfo{year}{1988}).

\bibitem[{\citenamefont{Romaides et~al.}(1994)\citenamefont{Romaides, Sands,
  Eckhardt, Fischbach, Talmadge, and Kloor}}]{Romaides1994}
\bibinfo{author}{\bibfnamefont{A.~J.} \bibnamefont{Romaides}},
  \bibinfo{author}{\bibfnamefont{R.~W.} \bibnamefont{Sands}},
  \bibinfo{author}{\bibfnamefont{D.~H.} \bibnamefont{Eckhardt}},
  \bibinfo{author}{\bibfnamefont{E.}~\bibnamefont{Fischbach}},
  \bibinfo{author}{\bibfnamefont{C.~L.} \bibnamefont{Talmadge}},
  \bibnamefont{and} \bibinfo{author}{\bibfnamefont{H.~T.} \bibnamefont{Kloor}},
  \bibinfo{journal}{Phys. Rev. D} \textbf{\bibinfo{volume}{50}}
  (\bibinfo{year}{1994}).

\bibitem[{\citenamefont{Taylor}(1993)}]{Taylor1993}
\bibinfo{author}{\bibfnamefont{J.~H.} \bibnamefont{Taylor}},
  \bibinfo{journal}{Class. Quantum Grav.} \textbf{\bibinfo{volume}{10}}
  (\bibinfo{year}{1993}).

\bibitem[{\citenamefont{Stairs et~al.}(1998)\citenamefont{Stairs, Arzoumanian,
  Camilo, Lyne, Nice, Taylor, Thorsett, and Wolszczan}}]{Stairs1998}
\bibinfo{author}{\bibfnamefont{I.~H.} \bibnamefont{Stairs}},
  \bibinfo{author}{\bibfnamefont{Z.}~\bibnamefont{Arzoumanian}},
  \bibinfo{author}{\bibfnamefont{F.}~\bibnamefont{Camilo}},
  \bibinfo{author}{\bibfnamefont{A.~G.} \bibnamefont{Lyne}},
  \bibinfo{author}{\bibfnamefont{D.~J.} \bibnamefont{Nice}},
  \bibinfo{author}{\bibfnamefont{J.~H.} \bibnamefont{Taylor}},
  \bibinfo{author}{\bibfnamefont{S.~E.} \bibnamefont{Thorsett}},
  \bibnamefont{and}
  \bibinfo{author}{\bibfnamefont{A.}~\bibnamefont{Wolszczan}},
  \bibinfo{journal}{ApJ} \textbf{\bibinfo{volume}{505}} (\bibinfo{year}{1998}).

\bibitem[{\citenamefont{Damour and Schaefer}(1991)}]{Damour1991}
\bibinfo{author}{\bibfnamefont{T.}~\bibnamefont{Damour}} \bibnamefont{and}
  \bibinfo{author}{\bibfnamefont{G.}~\bibnamefont{Schaefer}},
  \bibinfo{journal}{Phys. Rev. Lett.} \textbf{\bibinfo{volume}{66}}
  (\bibinfo{year}{1991}).

\bibitem[{\citenamefont{Barack et~al.}(2010)\citenamefont{Barack, Damour, and
  Sago}}]{Barack2010}
\bibinfo{author}{\bibfnamefont{L.}~\bibnamefont{Barack}},
  \bibinfo{author}{\bibfnamefont{T.}~\bibnamefont{Damour}}, \bibnamefont{and}
  \bibinfo{author}{\bibfnamefont{N.}~\bibnamefont{Sago}},
  \bibinfo{journal}{Phys. Rev. D} \textbf{\bibinfo{volume}{82}}
  (\bibinfo{year}{2010}), \eprint{1008.0935}.

\bibitem[{\citenamefont{Ori and Thorne}(2000)}]{Ori2000}
\bibinfo{author}{\bibfnamefont{A.}~\bibnamefont{Ori}} \bibnamefont{and}
  \bibinfo{author}{\bibfnamefont{K.~S.} \bibnamefont{Thorne}},
  \bibinfo{journal}{Phys. Rev. D} \textbf{\bibinfo{volume}{62}}
  (\bibinfo{year}{2000}), \eprint{0003032}.

\bibitem[{\citenamefont{Barack and Sago}(2010)}]{BarackSago2010}
\bibinfo{author}{\bibfnamefont{L.}~\bibnamefont{Barack}} \bibnamefont{and}
  \bibinfo{author}{\bibfnamefont{N.}~\bibnamefont{Sago}},
  \bibinfo{journal}{Phys. Rev. D} \textbf{\bibinfo{volume}{81}}
  (\bibinfo{year}{2010}), \eprint{1002.2386}.

\bibitem[{\citenamefont{Sago et~al.}(2008)\citenamefont{Sago, Barack, and
  Detweiler}}]{Sago2008}
\bibinfo{author}{\bibfnamefont{N.}~\bibnamefont{Sago}},
  \bibinfo{author}{\bibfnamefont{L.}~\bibnamefont{Barack}}, \bibnamefont{and}
  \bibinfo{author}{\bibfnamefont{S.}~\bibnamefont{Detweiler}},
  \bibinfo{journal}{Phys. Rev. D} \textbf{\bibinfo{volume}{78}}
  (\bibinfo{year}{2008}), \eprint{0810.2530}.

\bibitem[{\citenamefont{Barack and Lousto}(2005)}]{Barack2005}
\bibinfo{author}{\bibfnamefont{L.}~\bibnamefont{Barack}} \bibnamefont{and}
  \bibinfo{author}{\bibfnamefont{C.~O.} \bibnamefont{Lousto}},
  \bibinfo{journal}{Phys. Rev. D} \textbf{\bibinfo{volume}{72}}
  (\bibinfo{year}{2005}).

\bibitem[{\citenamefont{Dolan et~al.}(2014)\citenamefont{Dolan, Warburton,
  Harte, Le~Tiec, Wardell, and Barack}}]{Dolan2014}
\bibinfo{author}{\bibfnamefont{S.~R.} \bibnamefont{Dolan}},
  \bibinfo{author}{\bibfnamefont{N.}~\bibnamefont{Warburton}},
  \bibinfo{author}{\bibfnamefont{A.~I.} \bibnamefont{Harte}},
  \bibinfo{author}{\bibfnamefont{A.}~\bibnamefont{Le~Tiec}},
  \bibinfo{author}{\bibfnamefont{B.}~\bibnamefont{Wardell}}, \bibnamefont{and}
  \bibinfo{author}{\bibfnamefont{L.}~\bibnamefont{Barack}},
  \bibinfo{journal}{Phys. Rev. D} \textbf{\bibinfo{volume}{89}}
  (\bibinfo{year}{2014}), \eprint{1312.0775}.

\bibitem[{\citenamefont{Barack}(2009)}]{Barack2009}
\bibinfo{author}{\bibfnamefont{L.}~\bibnamefont{Barack}},
  \bibinfo{journal}{Class. Quant. Grav.} \textbf{\bibinfo{volume}{26}}
  (\bibinfo{year}{2009}), \eprint{0908.1664}.

\bibitem[{\citenamefont{Sotiriou et~al.}(2008)\citenamefont{Sotiriou, Faraoni,
  and Liberati}}]{Sotiriou2008}
\bibinfo{author}{\bibfnamefont{T.~P.} \bibnamefont{Sotiriou}},
  \bibinfo{author}{\bibfnamefont{V.}~\bibnamefont{Faraoni}}, \bibnamefont{and}
  \bibinfo{author}{\bibfnamefont{S.}~\bibnamefont{Liberati}},
  \bibinfo{journal}{Int. J. Mod. Phys. D} \textbf{\bibinfo{volume}{17}}
  (\bibinfo{year}{2008}), \eprint{0707.2748}.

\bibitem[{\citenamefont{Moore et~al.}(2015)\citenamefont{Moore, Cole, and
  Berry}}]{Moore2015}
\bibinfo{author}{\bibfnamefont{C.~J.} \bibnamefont{Moore}},
  \bibinfo{author}{\bibfnamefont{R.~H.} \bibnamefont{Cole}}, \bibnamefont{and}
  \bibinfo{author}{\bibfnamefont{C.~P.~L.} \bibnamefont{Berry}},
  \bibinfo{journal}{Class. Quant. Grav.} \textbf{\bibinfo{volume}{32}}
  (\bibinfo{year}{2015}), \eprint{1408.0740}.

\bibitem[{\citenamefont{Gair et~al.}(2004)\citenamefont{Gair, Barack,
  Creighton, Cutler, Larson, Phinney, and Vallisneri}}]{Gair2004}
\bibinfo{author}{\bibfnamefont{J.~R.} \bibnamefont{Gair}},
  \bibinfo{author}{\bibfnamefont{L.}~\bibnamefont{Barack}},
  \bibinfo{author}{\bibfnamefont{T.}~\bibnamefont{Creighton}},
  \bibinfo{author}{\bibfnamefont{C.}~\bibnamefont{Cutler}},
  \bibinfo{author}{\bibfnamefont{S.~L.} \bibnamefont{Larson}},
  \bibinfo{author}{\bibfnamefont{E.~S.} \bibnamefont{Phinney}},
  \bibnamefont{and}
  \bibinfo{author}{\bibfnamefont{M.}~\bibnamefont{Vallisneri}},
  \bibinfo{journal}{Class. Quant. Grav.} \textbf{\bibinfo{volume}{21}}
  (\bibinfo{year}{2004}), \eprint{0405137}.

\bibitem[{\citenamefont{Maggiore}(2007)}]{Maggiore}
\bibinfo{author}{\bibfnamefont{M.}~\bibnamefont{Maggiore}},
  \emph{\bibinfo{title}{Gravitational Waves, Vol. 1: Theory and Experiments}}
  (\bibinfo{publisher}{Oxford University Press}, \bibinfo{year}{2007}).

\bibitem[{\citenamefont{Babak et~al.}(2017)\citenamefont{Babak, Gair, Sesana,
  Barausse, Sopuerta, Berry, Berti, Amaro-Seoane, Petiteau, and
  Klein}}]{Babak2017}
\bibinfo{author}{\bibfnamefont{S.}~\bibnamefont{Babak}},
  \bibinfo{author}{\bibfnamefont{J.}~\bibnamefont{Gair}},
  \bibinfo{author}{\bibfnamefont{A.}~\bibnamefont{Sesana}},
  \bibinfo{author}{\bibfnamefont{E.}~\bibnamefont{Barausse}},
  \bibinfo{author}{\bibfnamefont{C.~F.} \bibnamefont{Sopuerta}},
  \bibinfo{author}{\bibfnamefont{C.~P.~L.} \bibnamefont{Berry}},
  \bibinfo{author}{\bibfnamefont{E.}~\bibnamefont{Berti}},
  \bibinfo{author}{\bibfnamefont{P.}~\bibnamefont{Amaro-Seoane}},
  \bibinfo{author}{\bibfnamefont{A.}~\bibnamefont{Petiteau}}, \bibnamefont{and}
  \bibinfo{author}{\bibfnamefont{A.}~\bibnamefont{Klein}},
  \bibinfo{journal}{Phys. Rev. D} \textbf{\bibinfo{volume}{95}}
  (\bibinfo{year}{2017}), \eprint{1703.09722}.

\bibitem[{\citenamefont{Robson et~al.}(2018)\citenamefont{Robson, Cornish, and
  Liu}}]{Robson2018}
\bibinfo{author}{\bibfnamefont{T.}~\bibnamefont{Robson}},
  \bibinfo{author}{\bibfnamefont{N.}~\bibnamefont{Cornish}}, \bibnamefont{and}
  \bibinfo{author}{\bibfnamefont{C.}~\bibnamefont{Liu}} (\bibinfo{year}{2018}),
  \eprint{1803.01944}.

\bibitem[{\citenamefont{Mirshekari and Will}(2013)}]{Mirshekari2013}
\bibinfo{author}{\bibfnamefont{S.}~\bibnamefont{Mirshekari}} \bibnamefont{and}
  \bibinfo{author}{\bibfnamefont{C.~M.} \bibnamefont{Will}},
  \bibinfo{journal}{Phys. Rev. D} \textbf{\bibinfo{volume}{87}}
  (\bibinfo{year}{2013}), \eprint{1301.4680}.

\bibitem[{\citenamefont{De~Laurentis et~al.}(2016)\citenamefont{De~Laurentis,
  Porth, Bovard, Ahmedov, and Abdujabbarov}}]{DeLaurentis2016}
\bibinfo{author}{\bibfnamefont{M.}~\bibnamefont{De~Laurentis}},
  \bibinfo{author}{\bibfnamefont{O.}~\bibnamefont{Porth}},
  \bibinfo{author}{\bibfnamefont{L.}~\bibnamefont{Bovard}},
  \bibinfo{author}{\bibfnamefont{B.}~\bibnamefont{Ahmedov}}, \bibnamefont{and}
  \bibinfo{author}{\bibfnamefont{A.}~\bibnamefont{Abdujabbarov}},
  \bibinfo{journal}{Phys. Rev. D} \textbf{\bibinfo{volume}{94}}
  (\bibinfo{year}{2016}), \eprint{1611.05766}.

\bibitem[{\citenamefont{Huerta et~al.}(2017)\citenamefont{Huerta, Kumar,
  Agarwal, George, Schive, Pfeiffer, Haas, Ren, Chu, Boyle
  et~al.}}]{Huerta2017}
\bibinfo{author}{\bibfnamefont{E.~A.} \bibnamefont{Huerta}},
  \bibinfo{author}{\bibfnamefont{P.}~\bibnamefont{Kumar}},
  \bibinfo{author}{\bibfnamefont{B.}~\bibnamefont{Agarwal}},
  \bibinfo{author}{\bibfnamefont{D.}~\bibnamefont{George}},
  \bibinfo{author}{\bibfnamefont{H.-Y.} \bibnamefont{Schive}},
  \bibinfo{author}{\bibfnamefont{H.~P.} \bibnamefont{Pfeiffer}},
  \bibinfo{author}{\bibfnamefont{R.}~\bibnamefont{Haas}},
  \bibinfo{author}{\bibfnamefont{W.}~\bibnamefont{Ren}},
  \bibinfo{author}{\bibfnamefont{T.}~\bibnamefont{Chu}},
  \bibinfo{author}{\bibfnamefont{M.}~\bibnamefont{Boyle}},
  \bibnamefont{et~al.}, \bibinfo{journal}{Phys. Rev. D}
  \textbf{\bibinfo{volume}{95}} (\bibinfo{year}{2017}), \eprint{1609.05933}.

\end{thebibliography}
\end{document}